\documentclass[twocolumn,prl,amsmath,amssymb,superscriptaddress,showpacs]{revtex4}

\usepackage{graphicx}
\usepackage{dcolumn}
\usepackage{bm}
\usepackage{times}

\begin{document}


\title{Three dimensional super-resolution in metamaterial slab lenses}

\author{F. Mesa}
\affiliation{Dept.~de F\'isica Aplicada I. Universidad de Sevilla,
41012-Sevilla (Spain)}
\author{R. Marqu\'es}
\affiliation{Dept.~de Electr\'onica y Electromagnetismo.
Universidad de Sevilla, 41012-Sevilla (Spain)}
\author{M. Freire}
\affiliation{Dept.~de Electr\'onica y Electromagnetismo.
Universidad de Sevilla, 41012-Sevilla (Spain)}
\author{J. D. Baena}
\affiliation{Dept.~de Electr\'onica y Electromagnetismo.
Universidad de Sevilla, 41012-Sevilla (Spain)}

\date{\today}

\begin{abstract}

This letter presents a theoretical and experimental study on the
viability of obtaining three dimensional super-resolution (i.e.
resolution overcoming the diffraction limit for all directions in
space) by means of metamaterial slab lenses. Although the source
field cannot be actually reproduced at the back side of the lens
with super-resolution in all space directions, the matching
capabilities of metamaterial slabs does make it possible the
detection of images with three-dimensional super-resolution. This
imaging takes place because of the coupling between the evanescent
space harmonic components of the field generated at both the
source and the detector.

\end{abstract}


\maketitle

It is well known from the early works of Veselago \cite{Veselago1} that a
slab made of a left-handed medium will focus the electromagnetic energy
coming from a point source to another point located at the opposite side of the
slab. An experimental confirmation of this focusing of energy has been reported
in  \cite{Houck}. Subsequent works \cite{Pendry,Grbic,Lagarkov} have shown that,
under some circumstances, metamaterial lenses can produce images at certain
planes with a resolution beyond the classical diffraction limit, or ``super
resolution imaging'' (SRI). This SRI has been attributed to an amplification,
inside the lens, of the evanescent Space Fourier Harmonics (SFHs) coming from
the source \cite{Pendry}. In \cite{Fang} it has been also discussed that this
process gives rise to fields that decay exponentially from the lens towards the
image, which causes  super-resolution to take place only in \emph{planes}
parallel to the slab interfaces. In the direction perpendicular to the lens, a
strong decay of the field is observed, and thus a three dimensional (3D)
picture of the source cannot be obtained from the field pattern at the back
side of the lens. In other words, super-resolution in the transverse
directions is obtained at the price of a drastic loss of resolution in the
longitudinal direction. This fact has been corroborated by the
field measurements recently reported  in \cite{Grbic}, where the field growth
from the image plane to the ``super lens'' can be clearly appreciated. Other
experimental results also lead to the same conclussion, showing that images ``of finite
depth'' \cite{Lagarkov} cannot be directly obtained from field measurements in SRI
experiments. Although the above facts seem to be well stablished, new
experimental results recently reported by some of the authors
\cite{Freire} have suggested the possibility of also obtaining super-resolution
in the longitudinal direction or, in other words, three-dimensional
super-resolution imaging (3D-SRI). In these experiments, a 3D map of a
point-like source (i.e. a source of sub-wavelength size) was obtained at the
image side of the lens. The aim of the present letter is then to provide the
general theory underlying this 3D-SRI. It will be shown that 3D-SRI of
point-like sources is actually a general property of metamaterial slabs when the
appropriate detection procedure is followed.

The main difference between SRI and more conventional imaging processes is that, in the
former one, the information for the image formation is carried out by \emph{evanescent}
SFHs \cite{Pendry} whereas in  conventional imaging the information is carried out by
the \emph{propagative} SFHs. Since  evanescent fields cannot carry power, and
any
image measurement requires some power  transmission, it is apparent that the
dectection of a super-resolution image should
substantially affect the fields around the detector. This intrinsic perturbation
of the fields in a super-resolution image detection resembles the problem of the
tunneling effect. As is well known, tunneling of power is due to the excitation of
a pair of evanescent electromagnetic waves, whose interference gives rise to a
non-vanishing flux of power. ``Perfect tunneling'' of power in a waveguide filled by a
metamaterial has been recently reported by some of the authors in \cite{Baena}. In
this work it is shown that maximum tunneling of power in a setup with identical input
and  output waveguides is achieved when the output is placed at a distance from the input
equal to that from the source to the image in a metamaterial super-lens. This suggests
that a similar effect could take place in a metamaterial super-lens, provided
that a  detector identical to the source is used for the measurements. In the
following it  will be shown how we can take advantage of such effect in order to
obtain 3D-SRI with metamaterial slabs.

\begin{figure}[ht]
  \centering
  \includegraphics[angle=-90,width=0.8\columnwidth]{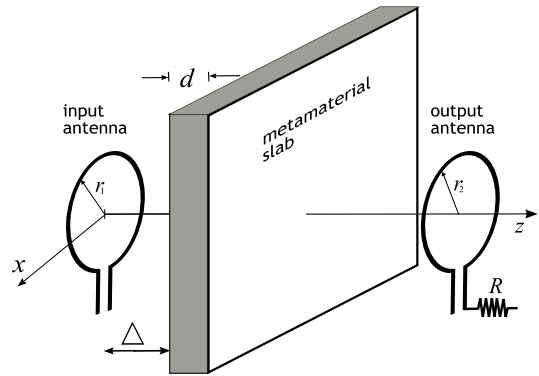}
  \caption{\label{lens} Geometry of a metamaterial lens. The source
  is a loop antenna located at $z=0$.
  The detector is also an identical loop antenna. The lens is formed
  either by a left-handed metamaterial slab or by another planar
  device that produces SRI through amplification
  of evanescent harmonics. The output antenna can eventually be loaded
  with microwave resistors in order to minimize field perturbation.}
\end{figure}

The present study starts with the canonical problem of the formation of images by a
left-handed slab of thickness $d$ characterized by $\epsilon /\epsilon_0=\mu
/\mu_0=-1+i\delta$, where $\delta \ll 1$ accounts for the necessary losses factor to
avoid the divergence of field integrals \cite{Garcia,SmithAPL03,MarquesMOTL}. In our
study, and following an usual procedure in microwave SRI experiments, the \emph{source}
will be an antenna (specifically, a loop antenna) whose plane is located
parallel to the slab interfaces, as is shown in Fig.~\ref{lens}. The field beyond
the lens will be scanned by an output antenna that plays the role of
\emph{detector}. The source here employed is equivalent to an homogeneous surface
distribution of magnetic dipoles inside the loop given by
$M_s=I_0\mathrm{e}^{-i\omega t}$, where $I$ denotes the amplitude of the
imposed time-harmonic current in the loop. The computation of the longitudinal
magnetic field, $H_z$, at the image side of the slab ($z>\Delta + d$) can be carried out
by means of the following double inverse Fourier transform:
\begin{multline}\label{eq:Hz}
   H_z(x,y,z) = \dfrac{1}{4\pi^2} \iint \mathrm{d}k_x\mathrm{d} k_y \\
  \widetilde{G}(k_x,k_y;z)\tilde{M}_s(k_x,k_y)\,\mathrm{e}^{i(k_xx+k_yy)}\;,
\end{multline}
where $\widetilde{G}(k_x,k_y;z)$ and $\widetilde{M}_s(k_x,k_y;z)$ are
respectively the Fourier transforms of the Green's function of the structure
under study and of the spatial surface distribution of magnetic dipoles. After
applying the duality principle to expression (6) in \cite{MarquesMOTL} and
taking $\Delta=d/2$, the Fourier-transform of the Green's function is found to
be
\begin{equation}\label{eq:G}
   \widetilde{G}(k_x,k_y;z) =
     \dfrac{-k_t^2 2\beta\mathrm{e}^{-i\beta_0(z+d)}}
      {\omega\left\{ (\beta\mu_0-\beta_0\mu)^2\mathrm{e}^{i\beta d}
      - (\beta\mu_0+\beta_0\mu)^2\mathrm{e}^{-i\beta d}\right\} } \;,
\end{equation}
with $k_t^2 = (k_x^2+k_y^2)$ and $\beta = \sqrt{\omega^2\varepsilon\mu -
k_t^2}$, $\beta_0 = \sqrt{\omega^2\varepsilon_0\mu_0 - k_t^2}$.
Since $E_z=0$ in the present case, the remaining field components can be all
deduced from $H_z$ \cite{MarquesMOTL}.

\begin{figure}[h]
\hspace*{-6mm}\includegraphics[angle=-90,width=1.1\columnwidth]{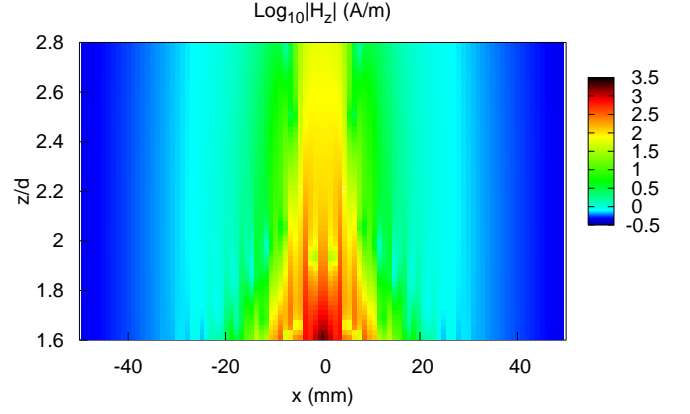}
  \caption{\label{H-field} %
  Map of $\log_{10}|H_z|$ corresponding to the field generated at the  back
  side  of the lens by a source loop antenna  shown in  Fig.~\ref{lens}. The
  slab is made of a left-handed medium of  thickness $d=4\,$mm with
  $\epsilon/\epsilon_0=\mu /\mu_0=-1+i0.001$, and is  separated a
  distance $\Delta=d/2$ from the source. The source is considered, for
  simplicity, a  \emph{circular} loop antenna of radius $r_1=5\,$mm,. The
  wire radius is of $0.2\,$mm. The operation frequency is 3 GHz.}
\end{figure}
\begin{figure}[ht]
  \centering
  \includegraphics[angle=-90,width=0.90\columnwidth]{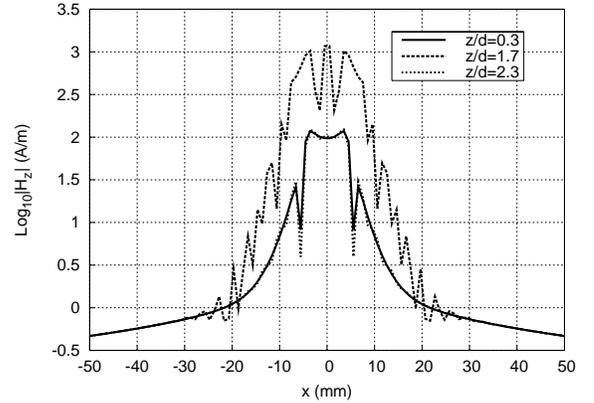}
  \caption{\label{H-plot} Plot of the $\log_{10}|H_z|$ at
  different $z$-planes of Fig.1. The slab and the source loop antenna is
  as in Fig.~\ref{H-field}.}
\end{figure}

The numerical computation of \eqref{eq:Hz} provides the map depicted in
Fig.~\ref{H-field} for the magnitude of $H_z$ in the $(x-z)$ plane at the back side
of the lens. It should be noted that, given that the operation frequency is 3
GHz ($\lambda_0=100\,$mm) and $d$ and $\Delta$ are taken much less than the
free-space wavelength, the ray model approximation cannot be here employed to
obtain the fields in the considered region (in other words, we are dealing with
near fields and therefore in a SRI situation). According to previous
discussions, Fig.~\ref{H-field} shows that the field magnitude
has a strong decay along the $z$ direction, so that no information about the
location of the source in this direction can be extracted from the field pattern.
A more detailed picture of the field distribution at two planes $z =(2\pm \delta) d$
is depicted in Fig.~\ref{H-plot} (with $\delta$ set to 0.3), and compared with the
field amplitude near the source at $z=\pm\delta d$ (both results are identical).
\begin{figure}[ht]
  \centering
\hspace*{-6mm}\includegraphics[angle=-90,width=1.1\columnwidth]{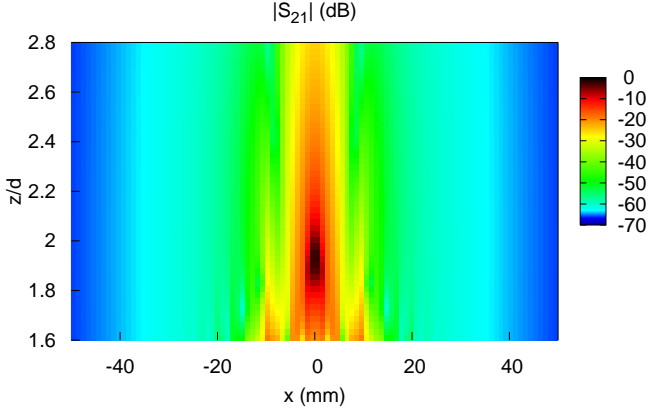}
  \caption{\label{t-map}%
  Map of the magnitude in decibels of the transmission coefficient,
$|S_{21}|$,  between the input and output antennas of Fig.~\ref{lens}. The
structural parameters are as in Fig.~\ref{H-field}. The output antenna is
identical to the input one.}
\end{figure}
As is expected from the properties of the metamaterial slab, the
field distribution at the $z= (2+\delta)d$ plane is almost
identical to that at the $z= \delta d$ plane, thus confirming that
the $z=\delta d$ and the $z=(2+\delta)d$ planes are in fact
\emph{conjugate} planes. On the contrary, the field distribution
at the $z=(2-\delta)d$ plane substantially differs in magnitude
from the field distribution at the $z=-\delta$ plane, near the
source. These facts corroborate the aforementioned discussions:
the super-resolution in the transverse directions (the lateral
dimensions of the source and the image are of about one tenth of
the wavelength) is compensated by an almost complete loss of
resolution in the longitudinal direction.

Next, the problem of the measurement of the image in this SRI
experiment will be considered. In the microwave range, the image
detection is performed by measuring the transmission coefficient
between a source antenna and a receiving antenna, which is scanned
in the image side of the lens \cite{Houck,Grbic,Lagarkov}. For
simplicity the receiving (or output) antenna will be assumed to be
identical to the input antenna employed as source. The
transmission coefficient is measured by connecting the input
antenna to a wave generator via a waveguide, and the output one to
a detector through another identical waveguide (more details of
this measurement setup are reported in \cite{Freire}). In this
approach, the metamaterial slab should be viewed as a
\emph{matching} device \cite{Veselago}, whose transmission
coefficient, $t$ (or $S_{21}$ in the usual microwave terminology),
is given by \cite{Pozar}
\begin{equation}\label{t}
t = \frac{2Z_{12}Z_0}{(Z_{11}+Z_0)(Z_{22}+Z_0)-Z_{12}^2} \;,
\end{equation}
where $Z_{ij}$ are the elements of the impedance matrix for the system formed by the
two antennas and the left-handed slab, and $Z_0$ is the characteristic impedance of
the input/output waveguide. In order to measure a super-resolution image, the size of
the loop antennas has to be smaller than the free space wavelength, which
additionally would make the real part of $Z_{ij}$ (the radiation resistance
\cite{Pozar}) be negligible with respect to its imaginary part; namely,
$Z_{ij}\simeq -i\omega L_{ij}$, where $L_{ij}$ is the inductance matrix of the
system. The diagonal terms of the inductance matrix correspond to the
inductances of a single loop antenna faced to the left-handed slab. However, in
the ``perfect lens'' configuration here considered, the slab does not affect the
fields around the source \cite{Pendry} and, therefore, $L_{11}=L_{22}\equiv L$,
where $L$ is the self-inductance of the loops in free space. The non-diagonal
terms of the inductance matrix account for the mutual inductance of the loop
antennas in the presence of the slab, namely $L_{12}=L_{21}\equiv M$.
Assuming that the wave\-guides have a low impedance (typically this
impedance is $Z_0=50\,\Omega$ for microwave measurements), it will be found
that $|Z_{ii}|>Z_0$. Thus, neglecting second order
terms in $Z_{ii}/Z_0$, the transmission coefficient in \eqref{t} can be
approximated as
\begin{equation}\label{tapprox}
t \approx \frac{2i\omega MZ_0}{\omega^2(L^2-M^2)+2i\omega L}\;.
\end{equation}
Note that the only spatial dependence in (\ref{tapprox}) comes from $M$ (since $L$
does not change with the position of the antennas), which causes that the
transmission
coefficient reaches its maximum ($|t|=1$) at those points where $M\to L$. Since both
the input and output antennas are identical, and the field at the source plane is
reproduced at the image plane (see Fig.~\ref{H-plot}), the condition $M\to L$ is expected
to be satisfied around the point $(x,y,z)=(0,0,2d)$. In consequence, a maximum of the
transmitted power should be detected at this particular point, which thus
appears as an ``effective focusing point'' of the perfect lens.

In order to show the above expected effects in a practical situation, the
magnitude of the transmission coefficient has been numerically computed for the
configuration under study. The  impedance matrix has been calculated by
imposing known currents, $I_i=1,0\,$A, in the loops, and then computing the
corresponding self and mutual inductances. The computation of the self
inductance is an standard electromagnetic problem, and the mutual inductance is
obtained after computing the flux of the magnetic field given by \eqref{eq:Hz}
across the surface of the output antenna. The transmission coefficient is
finally determined from (\ref{t}), assuming $Z_0=50\,\Omega$. In
Fig.~\ref{t-map} it is shown a map of the computed transmission coefficient for
a system composed of two identical lossless loop antennas. It can be
observed that this figure substantially differs from Fig.~\ref{H-field}; in
particular, a clear maximum of $|t|$ can be observed in Fig.~\ref{t-map} in the
neighborhood of the image at $(0,0,2d)$. These results clearly show the
difference between the transmitted power and the field distribution in the
absence of the output antenna, and also how 3D-SRI can be obtained when the
appropriate detector is used.

\begin{figure}[htb]
  \centering
  \includegraphics[angle=-90,width=0.9\columnwidth]{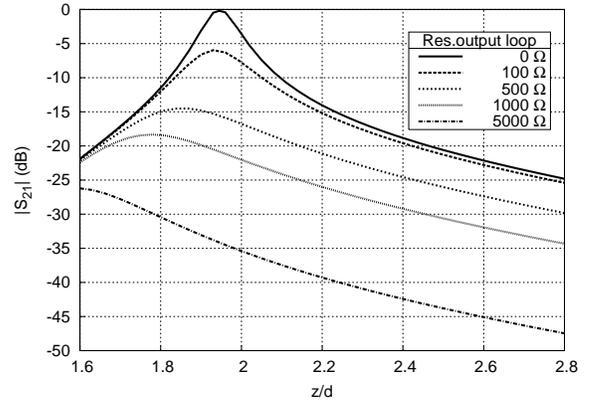}
  \caption{\label{R-t} %
  Magnitude of the computed transmission coefficient along the $Z$ axis between
  the input and output antennas of Fig.~\ref{lens} for different resistances
  loading the output antenna. The structural parameters are as in
  Fig.~\ref{H-field}.}
\end{figure}

Let us now consider the measurement of the unperturbed field. For this purpose, the
detector should be designed to affect the field distribution as less as possible. It
could be closely achieved by loading the output antenna with an additional high
resistance, which will significantly reduce the current induced at the output
antenna, as well as its generated field. In Fig.~\ref{R-t}
the computed values of the transmission coefficient along the
$z$-axis of Fig.~\ref{lens} are shown for different resistances,
$R$, loading the output antenna. As is expected, the curve for
$R=0$ shows a maximum near the location of the image, which thus
appears as a ``focusing point'' of the lens, at $z=2d$. (The small
difference between the actual location of this maximum and $z=2d$
can be attributed to the non negligible values of the ratio
$|Z_{ii}/Z_0| \approx 8$ in the problem under study). As is
expected, the curves for the highest values of $R$ resemble the
behavior of $|H_z|$ in Fig.~\ref{H-field} for the unperturbed
configuration.

\begin{figure}[ht]
  \centering
  \includegraphics[width=1.0\columnwidth]{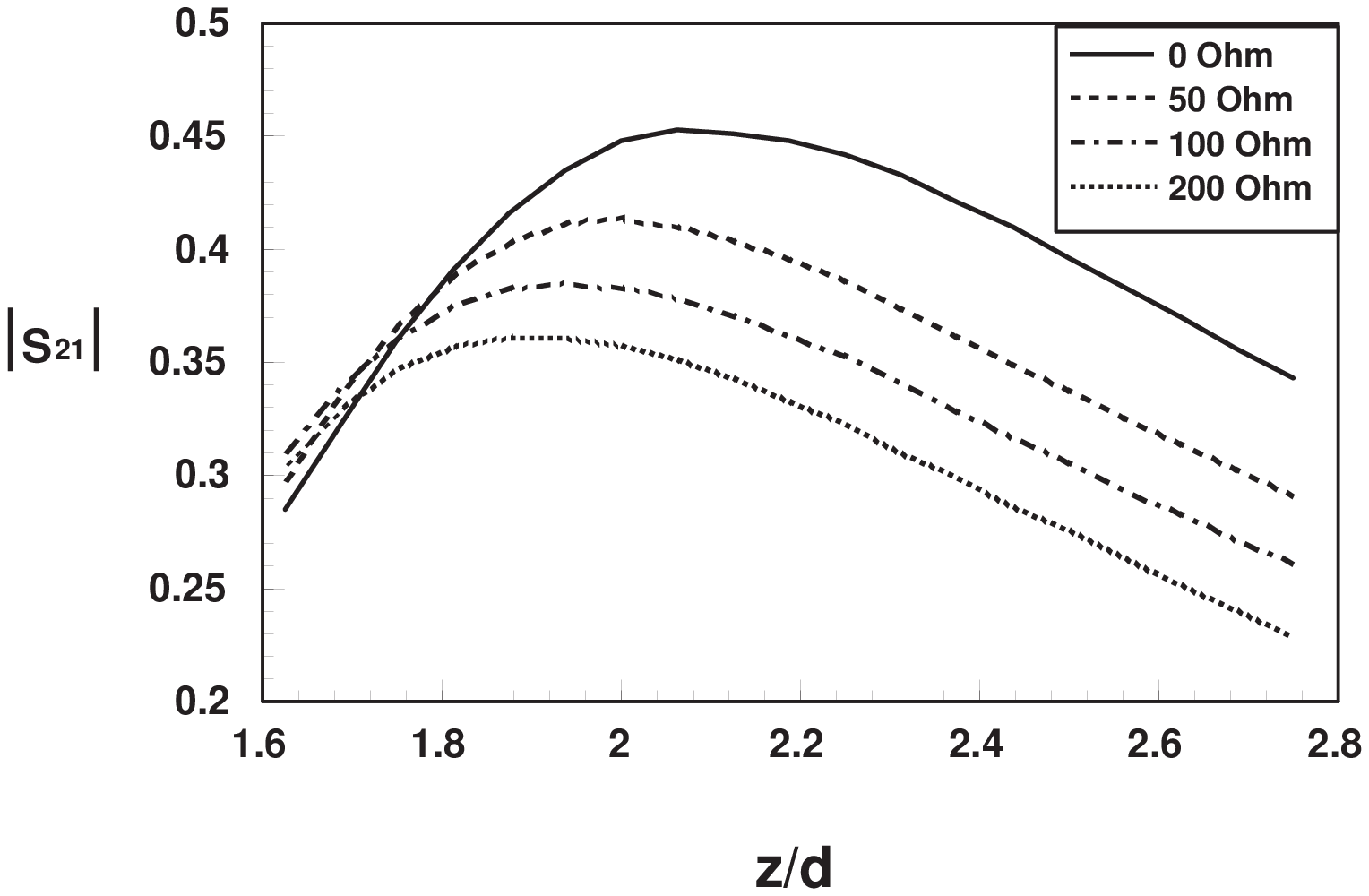}
  \caption{\label{R-exp} Plots of the measured magnitude of the
  transmission coefficient between the input and output antennas of
  Fig.~\ref{lens} when the slab is substituted by the magneto inductive lens
  reported in \cite{Freire}. (Freq=3GHz).}
\end{figure}

Let us now re-examine the experimental demonstration of 3D-SRI,
recently reported in \cite{Freire} at the light of the above
considerations. Although the underlying physics of the device
analyzed in \cite{Freire} is not exactly the same as that of the
left-handed perfect lens, both systems show the same process of
amplification of evanescent modes and are equivalent in practice
for the present purposes. Fig.~\ref{R-exp} shows the location of
the maximum of the transmission coefficient along the $z$ axis of
the \emph{magnetoinductive lens} reported in \cite{Freire}, for
different values of the resistance loading the output loop. The
experimental setup used to obtain these results is described in
detail in \cite{Freire}. The only difference is that, for
obtaining the results shown in Fig.~\ref{R-exp}, the output
antenna was loaded by different microwave resistors. A behavior
similar to that of Fig.~\ref{R-t} can be observed, thus confirming
the proposed theory.

In summary, SRI in metamaterial super-lenses has been
investigated. As is well known this imaging is primarily due to
the amplification of evanescent modes inside the lens. As a
consequence, super-resolution in the planes parallel to the slab
is unavoidably compensated by a drastic loss of resolution in the
direction perpendicular to the slab. However, since evanescent
modes do not carry power, any physical detection (namely, a
measurement) of the image at the back side of the lens will
require the existence of some amount of transmitted power from the
source to the detector, which can significantly affect the field
distribution. If this last effect is taken into account, the
secondary fields generated by the presence of the detector should
be considered. Following this approach, it has been shown that the
transmission coefficient between the source and the detector can
be made very high. Provided that the appropriate detector is
employed (typically a lossless output antenna identical to the
input antenna), this maximum will occur in a neighborhood of the
image, thus resulting in super-resolution imaging also in the
direction perpendicular to the slab. However, it should be
emphasized that, to obtain this 3D-SRI some previous knowledge of
the source is necessary in order to properly design the detector.
Thus, a general conclusion arises from the analysis:
\emph{Super-resolution in metamaterial super-lenses is always
incomplete: if the distance from the source to the lens is known,
the shape and characteristics of the source can be recovered
without uncertainty from the analysis of the field at the image
side of the lens. Conversely, if the shape and characteristics of
the source are known, it is possible to design an appropriate
detector in order to determine, without uncertainty, the source
location in three dimensional space from the transmission
coefficient between the source and the detector. However, it is
impossible to determine simultaneously, by means of a metamaterial
superlens, both the locations and the shape and characteristics of
an unknown source wit a resolution overcoming the diffraction
limit}. We feel that the reported analysis and experiments, as
well as the conclusions arising from them, will be of importance
in the design of metamaterial super-resolution devices.


\begin{thebibliography}
\expandafter\ifx\csname
natexlab\endcsname\relax\def\natexlab#1{#1}\fi
\expandafter\ifx\csname bibnamefont\endcsname\relax
  \def\bibnamefont#1{#1}\fi
\expandafter\ifx\csname bibfnamefont\endcsname\relax
  \def\bibfnamefont#1{#1}\fi
\expandafter\ifx\csname citenamefont\endcsname\relax
  \def\citenamefont#1{#1}\fi
\expandafter\ifx\csname url\endcsname\relax
  \def\url#1{\texttt{#1}}\fi
\expandafter\ifx\csname
urlprefix\endcsname\relax\def\urlprefix{URL }\fi
\providecommand{\bibinfo}[2]{#2}
\providecommand{\eprint}[2][]{\url{#2}}

\bibitem{Veselago1}
\bibinfo{author}{\bibfnamefont{V.~G.} \bibnamefont{Veselago}},
\bibinfo{journal}{Sov. Phys. USPEKHI} \textbf{\bibinfo{volume}{10}},
\bibinfo{pages}{509} (\bibinfo{year}{1968}).

\bibitem{Houck}
\bibinfo{author}{\bibfnamefont{A.A.} \bibnamefont{Houck}},
\bibinfo{author}{\bibfnamefont{J.B.} \bibnamefont{Brock}},
\bibinfo{author}{\bibfnamefont{I.L.} \bibnamefont{Chuang}},
\bibinfo{journal}{Phys. Rev. Lett.}
\textbf{\bibinfo{volume}{90}},
\bibinfo{pages}{137401} (\bibinfo{year}{2003}).

\bibitem{Pendry}
\bibinfo{author}{\bibfnamefont{J.~B.} \bibnamefont{Pendry}},
\bibinfo{journal}{Phys. Rev. Lett.}
\textbf{\bibinfo{volume}{85}},
\bibinfo{pages}{3966} (\bibinfo{year}{2000}).


\bibitem{Grbic}
\bibinfo{author}{\bibfnamefont{A.} \bibnamefont{Grbic}},
\bibinfo{author}{\bibfnamefont{G.~V.} \bibnamefont{Eleftheriades}},
\bibinfo{journal}{Phys. Rev. Lett.}
\textbf{\bibinfo{volume}{92}},
\bibinfo{pages}{117403} (\bibinfo{year}{2004}).

\bibitem{Lagarkov}
\bibinfo{author}{\bibfnamefont{A.~N.} \bibnamefont{Lagarkov}},
\bibinfo{author}{\bibfnamefont{V.~N.} \bibnamefont{Kissel}},
\bibinfo{journal}{Phys. Rev. Lett.}
\textbf{\bibinfo{volume}{92}},
\bibinfo{pages}{077401} (\bibinfo{year}{2004}).

\bibitem{Fang}
\bibinfo{author}{\bibfnamefont{N.} \bibnamefont{Fang}},
\bibinfo{author}{\bibfnamefont{X.} \bibnamefont{Zhang}},
\bibinfo{journal}{App. Phys. Lett.}
\textbf{\bibinfo{volume}{82}},
\bibinfo{pages}{161} (\bibinfo{year}{2003}).

\bibitem{Freire}
\bibinfo{author}{\bibfnamefont{M.} \bibnamefont{Freire}},
\bibinfo{author}{\bibfnamefont{R.} \bibnamefont{Marqu\'es}},
\bibinfo{journal}{App. Phys. Lett.}
\textbf{\bibinfo{volume}{86}},
\bibinfo{pages}{182505} (\bibinfo{year}{2005}).

\bibitem{Baena}
\bibinfo{author}{\bibfnamefont{J.D.} \bibnamefont{Baena}},
\bibinfo{author}{\bibfnamefont{L.} \bibnamefont{Jelinek}},
\bibinfo{author}{\bibfnamefont{F.} \bibnamefont{Marqu\'es}},
\bibinfo{author}{\bibfnamefont{F.} \bibnamefont{Medina}},
\bibinfo{journal}{Phys. Rev. B} (accepted)

\bibitem{Garcia}
\bibinfo{author}{\bibfnamefont{N.} \bibnamefont{Garc\'ia}},
\bibinfo{author}{\bibfnamefont{M.} \bibnamefont{Nieto-Vesperinas}},
\bibinfo{journal}{Phys. Rev. Lett.}
\textbf{\bibinfo{volume}{88}},
\bibinfo{pages}{207403} (\bibinfo{year}{2002}).

\bibitem{SmithAPL03}
\bibinfo{author}{\bibfnamefont{D.~R.} \bibnamefont{Smith}},
\bibinfo{author}{\bibfnamefont{D.} \bibnamefont{Schurig}},
\bibinfo{author}{\bibfnamefont{M.} \bibnamefont{Rosenbluth}},
\bibinfo{author}{\bibfnamefont{S.} \bibnamefont{Schultz}},
\bibinfo{author}{\bibfnamefont{S.} \bibnamefont{Anantha Ramakrishna}},
\bibinfo{author}{\bibfnamefont{J.~B.} \bibnamefont{Pendry}},
\bibinfo{journal}{App. Phys. Lett.}
\textbf{\bibinfo{volume}{82}},
\bibinfo{pages}{1506} (\bibinfo{year}{2003}).

\bibitem{MarquesMOTL}
\bibinfo{author}{\bibfnamefont{R.} \bibnamefont{Marqu\'es}},
\bibinfo{author}{\bibfnamefont{J.} \bibnamefont{Baena}},
\bibinfo{journal}{Microwave and Opt. Tech. Lett.}
\textbf{\bibinfo{volume}{41}},
\bibinfo{pages}{290} (\bibinfo{year}{2004}).

\bibitem{Veselago}
\bibinfo{author}{\bibfnamefont{V.~G.} \bibnamefont{Veselago}},
http://xxx.lanl.gov/ftp/cond-mat/papers/0501/0501438.pdf


\bibitem{Pozar}
\bibinfo{author}{\bibfnamefont{D.~M.} \bibnamefont{Pozar}},
\emph{\bibinfo{title}{Microwave Engineering}}
(\bibinfo{publisher}{J.Wiley \& Sons},
\bibinfo{address}{N.York},
\bibinfo{year}{1998}), \bibinfo{edition}{2nd} ed.


\end{thebibliography}
\end{document}